# Anisotropic, non-monotonic behavior of the superconducting critical current in thin YBa$_2$Cu$_3$O$_{7-\delta}$ films on vicinal SrTiO$_3$ surfaces

C. Cantoni, D. T. Verebelyi*, E. D. Specht, J. Budai, and D. K. Christen

*Oak Ridge National Laboratory, Oak Ridge, TN 37831*

* *American Superconductor Corporation, Westborough, MA 01581*

Contact Author:

Claudia Cantoni

Oak Ridge National Laboratory

Condensed Matter Sciences Division

P.O. Box 2008, MS-6061

Oak Ridge, TN 37831

Phone: (423) 574-6264, fax: (423) 574-6263

E-mail: cantonic@ornl.gov

ABSTRACT

The critical current density of epitaxial YBa$_2$Cu$_3$O$_{7-\delta}$ (YBCO) films grown on vicinal SrTiO$_3$ substrates was investigated by electrical transport measurements along and across the steps of the SrTiO$_3$ surface for a range of temperatures of 10 K to 85 K and in applied magnetic fields varying from 0 to 14 T. For vicinal angles of 4° and 8°, we found evidence of enhanced pinning in the longitudinal direction at low magnetic fields for a wide region of temperatures and attribute this phenomenon to anti-phase boundaries in the YBCO film. The transverse $J_c$ data showed a peak in the $J_c$(H) curve at low magnetic fields, which was explained on the basis of magnetic interaction between Abrikosov and Abrikosov-Josephson vortices. The in-plane $J_c$ anisotropy observed for vicinal angles of 0.4° was reversed with respect to the 8° and 4° samples. This phenomenon was interpreted on the basis of strain induced in the YBCO film by the stepped substrate's surface.

INTRODUCTION

In the last few years there has been much interest in studying the growth and the structural properties of YBa$_2$Cu$_3$O$_{7-\delta}$ (YBCO) films on stepped vicinal (001) SrTiO$_3$ (STO) surfaces. It has been found that steps induced on the substrate surface by a miscut angle $\psi$ between the *c*-axis and the surface normal, are responsible for a variety of peculiar phenomena in the YBCO films deposited on such substrates. These effects include modifications of the type of defects associated with the film lattice [1,2], anisotropy of the in-plane critical current density ($J_c$) [3-5], changes in the distribution of twin domains due to increased interfacial strain [6], vortex channeling effects [7], and change in the *c*-axis tilt of the YBCO film. Among others, Haage, and Jooss [1] have nicely documented the morphology of carefully reconstructed vicinal STO surfaces and the microstructure of the YBCO films grown of such surfaces. They focused on vicinal angles of ~ 10° towards the [100], which expose the (106) plane of STO, and reported significantly enhanced critical currents for conduction along the step ledges (longitudinal direction, or L). In their work $J_c$ is deduced from magnetization measurements at a temperature of 5 K and with an applied magnetic field that does not exceed 0.1 T.

The growth of YBCO films on vicinal surfaces is interesting for its implications on HTS-based 2nd generation wires, in which the YBCO film is deposited epitaxially on buffered <001>{100} biaxially textured metal tapes [8, 9]. In such metal tapes, each grain (10-100 µm wide) is misoriented by a few degrees with respect to the cube (001) orientation with the [100] parallel to the tape axis. In particular, each grain's *c*-axis is tilted from the tape surface normal by a different angle ranging between 0 and ~ 8°. In this sense the tape surface is naturally vicinal, and since texture is obtained by high-temperature anneal of the metal substrate, unit-cell-high steps are expected on the epitaxial oxide buffer layers deposited on them. Several studies have documented that when YBCO grows on a stepped STO surface, unit cells nucleating on adjacent terraces can be out of register because they are shifted vertically with respect to each other by a fraction of the YBCO unit cell [1, 10-14]. This situation induces an extended planar defect in the YBCO film which is referred to as antiphase boundary (APB). These defects will typically propagate until stacking faults are introduced in the YBCO lattice allowing matching of the basal lattice planes. When the surface projection of the substrate normal is carefully aligned towards the [010] and the vicinal angle $\psi$ is such that the terraces produced have a width $w$ ($w = a_{STO}/\tan \psi$) that is multiply commensurate with the STO lattice constant, then very straight, 1-unit-cell-high steps form in the STO surface. In this situation the APB's that form in the YBCO show a regular array and a structural width of less than 1 nm [1, 15]. This type of defect can act very differently from grain boundaries, which can have a larger structural width and a larger density of dislocations. Ch. Jooss et al. have shown that for miscuts of 9.46° along the [010], the superconducting critical current is not significantly suppressed for conduction in the direction perpendicular to the step ledges, for which the APB's constitute a barrier to the current flow [15]. In this case the effect of APB's on $J_c$ is different than that of grain boundaries of angle larger than 3°-4°, which strongly suppress the critical current [16, 17]. This may happen because the superconducting order parameter cannot be dramatically suppressed across APB's which have a structural with of only a fraction of the coherence length. In this sense APB's act like slightly weaker superconducting regions for transverse conduction and do not suppress $J_c$ significantly. At the same time the critical current for conduction parallel to the step edges is enhanced over typical values for a YBCO film on STO, showing that the interfaces formed by APB's are strong pinning sites for vortices. Although Jooss' analysis is very detailed, $J_c$ values are not obtained by direct measurement and are limited to a very narrow $T$-$H$ region. Depending on the specific application, HTS-based wires will have to operate at temperatures ranging from 30 to 77 K, and in magnetic fields ranging from tenths of a Tesla to several Tesla. In order to address the role of APB's boundaries in HTS coated conductors and compare to the effects of grain boundaries, more extensive transport $J_c$ measurements on prototype samples are needed.

EXPERIMENTAL

In this paper we present transport measurements of the critical current in the longitudinal (parallel to step ledges, L) and transverse (perpendicular to step ledges, T) directions, in a range of temperatures from 10 K to 85 K and in applied magnetic fields varying from 0 to 14 T. We analyzed YBCO films grown on STO substrates having a miscut angle of 0.4°, 4°, and 8° along the [010]. The substrates were characterized by x-ray diffraction (XRD), and atomic force microscopy (AFM) to accurately determine the direction of the substrate normal, the miscut angle, and to analyze the surface morphology. The YBCO films were analyzed by XRD and TEM to investigate the epitaxial relationship with the substrate, the defect microstructure, and the degree of twinning and strain in the lattice. Our results show that the APB's act as strong pinning centers in a wide region of temperatures for low magnetic fields, while the $J_c$ values measured in the two directions converge in magnetic fields greater than a few tenths of a Tesla. The transverse $J_c$ data showed a very peculiar field dependence, with a peak in the $J_c(H)$ curve at magnetic fields lower than the convergence point. This phenomenon can be interpreted on the basis of a model proposed by A. Gurevich and L. D. Cooley [18].

Unexpectedly, YBCO films deposited on substrates with relatively small miscut angle (0.4°) also showed anisotropy of the critical current, and these results are interpreted on the basis of strain induced in the YBCO film by the stepped substrate's surface.

All the YBCO films had a thickness of 150 to 250 *nm* and where grown under the same conditions using pulsed laser deposition with a wavelength of 248 *nm*, energy per pulse of 200 mJ, a substrate temperature of 780° C, and a background oxygen pressure of 100 mTorr. Typical YBCO films grown under these conditions on flat STO substrates, and with thickness ranging between 100 and 250 *nm*, showed $J_c$'s of about 4 MA/cm$^2$ at 77 K and zero applied magnetic field, and irreversibility fields at 77 K of 6-8 T.

For all samples the current transport measurement were performed using a 4 terminal configuration in which current is injected with alternating polarity through two Au-coated Cu pads pressed onto Ag sputtered contacts on the sample surface, and voltage is detected through separate pogo pins pressed onto two additional 4 mm-spaced Ag contacts. The pulsed current signal has a width less than 30 ms to prevent undesired Joule heating from power dissipation in the electrical contacts.

In the electrical transport measurements the longitudinal and transverse directions were individually probed by patterning the sample with a cross-conducting bridge using standard photolithographic technique. A schematic drawing of the type of bridge pattern used is shown in Fig. 1. Current was injected through the points A and B for the first measurement and through the points C and D for the second. In this way, conduction along each branch (100 μm wide) of the cross was obtained independently. Resistance vs. Temperature curves were measured for the two directions and both indicated the same critical temperature, as expected for a homogeneous YBCO film. For the YBCO film deposited on the 8° - miscut substrate $T_c$ was 93.3 K and the normal state resistivity in the transverse direction was 2.7 times larger than the longitudinal resistivity at T = 100 K. This value is consistent with partial conduction along the *c*-axis in a tilted film, according to the serial resistor model of ref. 1.

MORPHOLOGY AND ATOMIC STRUCTURE

Shown in Fig. 2 are the AFM micrographs of reconstructed STO surfaces with a miscut angle of 0.4° and 4° along the [010]. Surface reconstruction occurred during a 1 h-long anneal in flowing $O_2$ at T = 1000 °C. Line trace analysis showed that the step height was 1 STO unit cell in the case of 0.4°, and 2 or 3 STO unit cells for the 4° case. From the figure it is also evident that the steps of the 4° substrate are irregular along the longitudinal direction. The "non ideal" step morphology of this case is due to the fact that for a vicinal angle of 4° the terrace width $w$ is not a multiple of 3.905 Å, the lattice constant of STO. Therefore, the surface free energy is optimized by producing steps with heights equal to different integer multiples of $a_{STO}$. Imaging of the STO surface with $\psi$ = 8° by AFM was not possible since the terrace width (~28 Å) is smaller than the lateral resolution of the microscope. However, XRD measurements confirmed the direction of the substrate normal and yielded a value of 8.15° ± 0.02° for $\psi$. Such a vicinal angle is very close to the value 8.13° for which $w/a_{STO}$ = 7. Therefore, very straight steps with a height of 3.905 Å are expected in this case. An AFM image of the 120 nm-thick YBCO film grown on the 8°-miscut substrate is shown in Fig. 3. This figure shows that step-like features are retained in the YBCO morphology over a much coarser scale than that of the substrate's original morphology, indicating that many STO steps are overgrown by the YBCO film. Figure 4 is a high resolution TEM micrograph of a typical APB originated at the substrate step, which propagates for a few unit cells until the YBCO lattice planes match again as result of the introduction of a stacking fault. The micrograph was acquired with the zone axis parallel to step ledge direction <100>. This geometry allows viewing the sample along the transverse direction where substrate steps and the defects generated by them can be observed.

DISTRIBUTION OF TWIN DOMAINS AND LATTICE STRAIN

It is well known from experiments on untwinned single crystals that the normal-state transport properties of YBCO are anisotropic in the basal plane, and in particular the conductivity along *b* is larger than the conductivity along *a* [19]. However, YBCO films on STO are generally twinned with more or less equal population of twin domains, and such effect is not observed. Nevertheless, strong bias in the twin domain distribution has been observed in YBCO films on STO, and correlated to small deviation of the substrate normal from the [001] axis [6, 20]. The presence of steps in the substrate surface introduces anisotropic strain at the film/substrate interface because the terraces of the substrate can be more easily strained (by the nucleating YBCO film) perpendicular than parallel to the step edges. For this reason, it is important to measure the twin distribution in YBCO films on vicinal substrates, especially when analyzing anisotropy in transport properties. The distribution and population of the YBCO twin domains were determined using x-ray reciprocal space mapping, following the work of Brötz *et al*. [6]. We used a Cu Kα rotating anode source, a graphite monochromator, and a four-circle diffractometer. The incident beam size was 1 *mm* wide × 2 *mm* high, with 0.1° divergence. The diffracted beam was collimated by two slits, 1 *mm* wide by 25 *mm* high and 450 *mm* apart. Two-dimensional *hk* maps were taken of the YBCO (302) and (032) reflections, both with azimuth parallel to the ledges of the miscut substrate (STO [100]) and perpendicular to them (STO [010]). Precise measurements of the position and intensity of the peaks is made by least-squares fitting to the sum of four anisotropic, tilted Gaussians. Figure 5 shows a scan acquired for the 8°-miscut sample.

As also observed by Brötz *et al.*, the relative intensity of the (302) and (032) reflections in each direction differ due to a preferred orientation: because of the anisotropic strain at the interface, less of the material grows with the YBCO [100] (*a* axis) nearly parallel to the step ledges (we will call these D1 domains) than nearly perpendicular to them (D2 domains). However, Brötz *et al.* did not consider that the (302) and (032) reflections have different scattering factors, which also affect their intensities. To account for this effect, suppose the YBCO domains with *a*'s and *b*'s nearly parallel to the substrate ledges have population $f_a$ and $f_b$ respectively, and the scattering factors are $F_{302}$ and $F_{032}$. The intensities of the peaks with azimuth parallel to the ledges will then be proportional to:

$$I^{\parallel}_{302} = |F_{302}|^2 f_a$$
$$I^{\parallel}_{032} = |F_{032}|^2 f_b$$

Perpendicular to the ledges:

$$I^{\perp}_{302} = |F_{302}|^2 f_b$$
$$I^{\perp}_{032} = |F_{032}|^2 f_a$$

To find the preferred orientation, we have

$$f_b / f_a = \sqrt{I^{\parallel}_{032} I^{\perp}_{302} / I^{\parallel}_{302} I^{\perp}_{032}}$$

$$|F_{032}|^2 / |F_{302}|^2 = \sqrt{I^{\parallel}_{032} I^{\perp}_{032} / I^{\parallel}_{302} I^{\perp}_{302}}$$

We find, as do Brötz *et al.*, that the YBCO twin systems are aligned so that the <110> directions are rotated toward the ledge directions with respect to the substrate <110> directions. This means that the substrate terraces are tetragonally stretched by the YBCO film in the direction parallel to the step ledges and the angle $\zeta$ between YBCO [110] and YBCO [$\bar{1}$10] exceeds 90°. We find the angle $\zeta$ by noting that the (302) directions near the azimuth parallel to the step edges are separated by $\zeta/2-45°$ less than the (032) reflections; $\zeta$ calculated in this way is labeled $\zeta$[100]. Near the azimuth perpendicular to the step edges this is reversed; $\zeta$ calculated this way is labeled $\zeta$[010] and the two methods agree well. Shown in Table 1 are YBCO tilt, $f_b / f_a$, $|F_{032}|^2 / |F_{302}|^2$, and $\zeta$ for our 8°, 0.39°, and 0.42° samples. It is not clear whether the variation in the ratio $|F_{032}|^2 / |F_{302}|^2$ is due to structural differences or additional experimental uncertainty. Table 2 shows the in-plane YBCO lattice constants *a* and *b* of the same samples for each of the 4 twin domains.

Table I

| Substrate miscut | YBCO tilt | $f_b / f_a$ | $|F_{032}|^2 / |F_{302}|^2$ | $\zeta_{[100]}$ | $\zeta_{[010]}$ |
|---|---|---|---|---|---|
| 8.15 ± 0.02° | 8.50 ± 0.02° | 1.21 ± 0.05 | 0.86 ± 0.03 | 90.12 ± 0.01° | 90.11 ± 0.01° |
| 0.42 ± 0.02° | 0.02 ± 0.02° | 1.00 ± 0.04 | 0.77 ± 0.03 | 90.04 ± 0.01° | 90.04 ± 0.01° |
| 0.39 ± 0.02° | 0.03 ± 0.02° | 1.33 ± 0.05 | 0.88 ± 0.03 | 90.10 ± 0.01° | 90.06 ± 0.01° |

Table I: For each substrate vicinal angle the corresponding c-axis tilt of the YBCO film is shown. Other structural properties reported for each YBCO film are: ratio between different twin domains populations ($f_b/f_a$), ratio between scattering factors for YBCO (032) and YBCO (302) reflections ($|F_{032}|^2/|F_{302}|^2$), and angles between the YBCO <110> directions for the two twin systems as calculated from the peaks near both the substrate [100] azimuth ($\zeta_{[100]}$) and the substrate [010] azimuth ($\zeta_{[010]}$).

Table II

|  | Domain D1 | | Domain D2 | | | |
| --- | --- | --- | --- | --- | --- | --- |
| Substrate miscut | a // STO[100] (Å) | b // STO[010] (Å) | a // STO[010] (Å) | b // STO[100] (Å) | b/a D1 | b/a D2 |
| 8.15° | 3.8254 | 3.8889 | 3.8245 | 3.8898 | 1.0166 | 1.0171 |
| 0.42° | 3.8179 | 3.8747 | 3.8166 | 3.8772 | 1.0148 | 1.0159 |
| 0.39° | 3.8248 | 3.8722 | 3.8172 | 3.8811 | 1.0124 | 1.0167 |

Table II: in-plane lattice constants for YBCO films grown on STO substrates with different vicinal angles.

ELECTRICAL TRANSPORT MEASUREMENTS

We will start discussing the results obtained for the 8°-miscut substrate, which showed the most interesting features. In this case, the ratio between the populations of domains D2 and D1 is 1.21, and $\zeta = 90.12°$. Although the distribution of the twin domains is slightly asymmetric (55% of the material grows with the *b* axis oriented along the step ledges) this unbalance would not produce any visible anisotropy in the transport properties. In fact, a YBCO film grown on the 0.39°-miscut substrate with a ratio between twin populations of 1.33 did not show any asymmetry in the resistivity curves measured in the L and T directions. As shown in Table 2 for the 8°-miscut sample, the $a$ and $b$ lattice constants for the two twin variants are very close to each other and very similar to the YBCO bulk lattice constants ($a$ = 3.8227 Å; $b$ = 3.8872 Å). In this case, the YBCO film does not show any significant residual lattice strain. In addition, as illustrated in table 1, the YBCO film has grown with the $c$-axis almost parallel to the substrate $c$-axis. These considerations assure that if anisotropy is detected in the $J_c$ measurements this is not to be ascribed to lattice strain or twin domain distribution.

Shown in Fig 6a is a plot of the measured $J_c$ as function of applied magnetic field for several temperatures. Fig. 6b is an expanded view of Fig. 6a in the low field region with the addition of a data set obtained at T = 20 K. On a coarse field scale the curves acquired for transport along the L and T direction overlap for each temperature. However, on a finer scale remarkably different values of $J_c$ are obtained along the two directions for fields smaller than 0.3 T. For T = 20 K, $J_c$ at nominally zero field, measured along the step ledges, is 30 MA/cm$^2$, 4.8 times larger than the value obtained for transport perpendicular to the step ledges. The ratio $J_{c,L}/J_{c,T}$ decreases as the temperature increases but is greater than 1 for all temperatures. In the field regime 0 T ≤ H ≤ 0.3 T the effect of the defective microstructure is visible, especially at low temperatures, and a significant anisotropy is detected for transport in the basal plane. Figure 7 shows the dependence of $J_c$ on temperature for the two directions at H = 0 T, 0.05 T, 0.1 T, and 1 T, respectively. The dashed line represents the $J_c(T)$ curve in self field for a typical YBCO film of the same thickness on a non-vicinal substrate. At nominal zero applied field the critical current along the longitudinal direction is about 40% larger (at T=20 K) than the $J_c$ of YBCO films on flat substrates. In the orthogonal direction the critical current is suppressed by conduction across the planar defects as compared to that of typical YBCO films. However such suppression is not dramatic at high temperatures. For example, $J_{c,T}$ at 60 K is 4.7 MA/cm$^2$, roughly 32 % smaller than what expected for a regular YBCO film at the same temperature. Moreover, at even higher temperatures (T ≥ 70 K), the $J_c$ measured in the T direction is totally comparable to that of YBCO films on flat substrates. This is true not only in self field, as shown in Fig. 7, but at all applied fields. The temperature dependence of $J_c$ in zero applied field is also different for the T and L directions of the miscut sample, and the flat sample. In all cases $J_c(T)$ can be fitted by the expression $J_o(1-T/T_c)^x$ with $x = 1$ for L, $x = 0.5$ for T, and $x = 1.6$ for the YBCO film on flat substrate. However, already at H = 0.05 T both directions of the miscut sample exhibit the same temperature dependence with $x = 1$. At larger fields $x$ tends to increase, and at H=1T all three dependencies are equal with $x \sim 1.7$.

Fig. 6 shows $J_c(H)$ curves very similar to those of regular YBCO films on flat substrates. Therefore, one may conclude that in magnetic fields higher than a few tenths of a Tesla (that is for most wire power applications) the array of APB's present in the YBCO films does not affect the critical current density for either longitudinal or transverse transport. However, this general conclusion is subjected to some refinement when scrutinized in detail. Shown in Fig. 8 is a comparison between $J_c(H)$ curves for the L direction, the T direction and the flat sample for a wide range of magnetic fields at T= 40 K and T= 77 K. It is evident that at 77 K all three the curves overlap for magnetic field in excess of ~ 0.3 T and the flat-sample level of $J_c$ is equal to that of the T direction also for lower field values, suggesting a net enhancement of $J_c$ below 0.3 T for transport along the L direction as compared to a regular sample. This graph may lead us to conclude that the APB's are active flux pinning centers at low fields, and that at higher fields other types of defects present in both vicinal and flat YBCO films take over while the APB's become essentially transparent. However, this picture is not consistent with the lower-temperature observations. At 40 K, for example, although the L direction shows an enhancement in $J_c$ over the flat sample for very low fields (0 T<H<0.05 T), the flat sample shows a larger $J_c$ than either L or T directions in the range 0.05 T< H < 14 T. This may indicate that the flat sample has a larger density of certain flux-pinning defects, which are active in an intermediate range of fields and become important especially at low temperature, where the effects of thermal activation are minimal. Such defects would then lose effectiveness at higher fields, close to the irreversibility line where the $J_c$(H) curves of the flat and miscut samples appear to converge. It is conceivable that a lack of certain defects may occur in the miscut samples as a result of a different, more 2-dimensional growth mode generated by the presence of the steps. Nevertheless, our measurements indicate that, in the entire range of temperature investigated, the 8°-miscut sample in the L direction shows a higher $J_c$ than a regular sample in a certain range of small magnetic fields. We conclude that APB's act as strong flux pinning defects for parallel conduction only in a region of low magnetic fields. At the same time, because of their intrinsically small structural width, they do not dramatically suppress $J_c$ in perpendicular conduction. This is especially true above 60 K, where thermal activation and an increasing coherence length average out the local suppression of superconductivity by the APB's. At fields close to the irreversibility line the effect of APB's is negligible and YBCO films on vicinal surfaces behave similarly to films grown on flat surfaces.

YBCO films grown on 4° miscut substrates resembled in many aspects those grown on the 8°-miscut substrates. From Fig. 9, which shows the dependence of $J_c$ on field for one of such samples, it is evident that, as for the 8°-miscut sample, transport anisotropy is present only for very small magnetic fields. In the case of the 4°-miscut sample, the level of $J_c$ in the L direction is not larger than that measured on a regular YBCO film. This may be the result of a less than ideal substrate step morphology (see Fig. 1b), which might have originated larger defects in the YBCO, some of them possibly oriented perpendicular to the L direction.

NON-MONOTONIC FIELD-DEPENDENT $J_c$

Another signature of the presence of APB's is given by a non-monotonic behavior of $J_c$ as function of field in transverse conduction at low magnetic fields. In this region the $J_c(H)$ curves in the T direction show a peak that decreases in intensity and shifts to smaller magnetic fields as the temperature increases. This dependence is shown in Fig.10, which is a plot on a linear scale of most of the same data that was shown in Fig.6b. Although peaks in $J_c$ vs. field have been observed by other groups in some YBCO bi-crystals [21, 22], such a feature is very unusual and the majority of YBCO films grown by very different *in-situ* and *ex-situ* techniques show a monotonic decrease of the critical current with field. A peak in $J_c$ vs $H$ is sometimes observed as a result of pinning defects that are distributed in the superconductor with a characteristic spacing (the so called matching field or fish tail effect). However, if the distribution of APB's were to cause a matching field effect, we would expect no temperature dependence in the peak field, and would expect the peak in the L conduction branch, because it is in this direction that the APB's act as pinning centers. Moreover, the peak should appear at much larger fields than observed. In fact, the matching field $B_\phi$ is defined as $B_\phi = \Phi_o/a_o^2$ where $a_o$ is the spacing between defects. On the basis of the YBCO film morphology (see Fig. 2), we expect $a_o$ to be ~ 200 Å, and therefore $B_\phi$ ~ 5 T. Having excluded the matching field hypothesis, we may explain the peak on the basis of the interaction between Abrikosov (or A) vortices and so called Abrikosov-Josephson (or AJ) vortices located on the structural boundaries formed by the APB's, according to the model proposed in ref. 18. Such a model describes the behavior of vortices that reside on planar defects (perpendicular to the current flow) that do not show the weak-linked behavior of typical Josephson junctions. This is indeed the case of APB's, as well as of low-angle grain boundaries in HTS. According to the model, at planar defects, Abrikosov vortices transform into hybrid AJ vortices with a highly anisotropic Josephson core. The Josephson core has a size $l \approx (J_d/j_c) \cdot \xi$ along the defect and a width $\xi$ in the transverse direction. Here $J_d$ is the depairing current density, $j_c$ is the local barrier tunneling current density, and $\xi$ the superconducting coherence length. In general, the pinning potential $U(\mathbf{r})$ varies over the scale of $\xi$, and AJ vortices, with core size $l \gg \xi$ are much more weakly pinned than A vortices with a normal core of radius $\xi$. AJ vortices can easily slide along the planar defect giving rise to dissipation and therefore low critical current. For small enough magnetic fields, the longitudinal pinning force per unit length of vortex $f_{//}$ (for conduction across the barrier) can be written as:

$$f_{//} \cong \left(\frac{\Phi_o}{4\pi\lambda}\right)^2 \left(\frac{\alpha L}{L^2 + l^2} + \sqrt{H/\pi\Phi_o}\right); \tag{1}$$

Where $\Phi_o$ is the flux quantum, and $\lambda$ is the penetration depth [18]. The first term results from pinning of AJ vortices by inhomogeneities of the superconducting current along the direction of the planar interface. Such variations occur on a characteristic length scale L as result of structural and/or chemical variations in the barrier, and are described by the parameter $\alpha = \delta j_c/<j_c>$, with $\delta j_c$ amplitude of local supercurrent variations around the mean value $<j_c>$. The second term arises from magnetic interactions between the A vortices and the AJ vortices. The energy barriers caused by the field of the strongly pinned Abrikosov vortices pin the AJ vortices as they are driven along the planar defect, thereby impeding their motion. This pinning mechanism is strongly dependent on the magnitude of the field: if the magnetic field is too weak, vortices are very far apart and their interaction is negligible; if the field is too high the cores of the AJ vortices overlap and the phase coherence across the barrier is lost. In an intermediate range of fields $J_c$ can actually increase with $H$ because of the $\sqrt{H}$ dependence of the second term in Eq. 1. As the magnetic field approaches the value $H_d$ at which overlap of the AJ cores occurs, Eq.1 no longer holds and a more typical decrease of $J_c$ with field will be observed. Due to this cross-over between inter and intra grain pinning, a peak in $J_c$ is thus expected for conduction in the transverse direction for fields close to $H_d \cong (j_c/J_d)^2 \cdot H_{c2}$ [18]. In the case of our sample, an estimate of $H_d$ at low temperatures is given by $(1\times10^7/3\times10^8)^2\times120$ T = 0.13 T, close to the value $H = 0.11$ T for the peak field at 10 K, as shown in Fig.10. As the temperature increases, the field at which the AJ cores overlap decreases and this explains qualitatively the observed peak shift to lower fields with increasing temperature. Shown in Fig. 10 is also a best fit of the 10 K data using the relation $J_c = J_{co}\left(1 + \sqrt{H/H_s}\right)$ obtained by equating the Lorenz force $\Phi_o J_c/c$ to the pinning force $f_{//}$. $H_s$ is defined as

$$H_s = \frac{\pi \alpha^2 L^2 \Phi_o}{(L^2 + l^2)^2}$$

and represents the field above which $f_{//}$ is mostly determined by the shear between AJ and A vortices. From the best fit we obtain $H_s = 0.05$ T, smaller than $H_d$ as expected.

## $J_c$ ANISOTROPY IN NON-VICINAL YBCO FILMS

As mentioned in the introduction, anisotropy in $J_c$ was also observed, with the same type of measurements described above, on YBCO films grown on relatively flat substrates, having a miscut of only 0.4°. In this case, however, the sign of the anisotropy was reversed: larger critical currents were measured for conduction across the step ledges as compared to conduction along the step ledges. The most dramatic effect was observed on the 0.39° sample in Table 2 which showed a difference between $J_{c,L}$ and $J_{c,T}$ at 77 K of 3 MA/cm$^2$ (See Fig. 11a). For the 0.42° sample the anisotropy was still observed but became significant only at lower temperatures, as shown in Fig. 11b. According to the results shown in Table 1, in these samples the YBCO c-axis is aligned with the substrate's surface normal, and there is no significant imbalance in the population of twin domains. In addition, normal resistivity curves measured along the T and L direction overlapped. Table 2 indicates that in both the 0.42° and the 0.39° samples the YBCO in-plane lattice constants differ significantly from the bulk values. In particular, for both samples *a* and *b* lattice constants are smaller than the corresponding bulk values. The only exception was found for the *a*'s in domain D1 (*a*-axis parallel to step ledge) for the 0.39° sample, which are very close to the bulk a value (only 0.05% larger). The largest compression is measured in both samples for the *b*'s of the D1 domain, that is the *b*'s oriented perpendicular to the step ledges. For example, in the sample with $\psi = 0.39°$, the *b*'s perpendicular to the step ledges are 0.39% smaller than the bulk *b* lattice constant, while *a*'s and *b*'s in the D2 domain are 0.14% and 0.16% smaller than the bulk values, respectively. Studies on untwinned YBCO single crystals have shown that the pressure derivatives $dT_c/dp_{a,b}$ are large and opposite in sign for compression in the *a-b* plane, with compression along *b* leading to an increase in $T_c$ [23]. Although the effects of uniaxial stress on critical current density in YBCO are not known, on the basis of such findings we may surmise that compression in the direction of the YBCO chains leads to larger $J_c$ in this direction. Therefore, the larger critical current measured in the T direction would result from the larger compression experienced by the *b* axis for the twin domains having *b* perpendicular to the step ledges.

CONCLUSIONS

We showed that the steps formed on reconstructed STO surfaces as result of a miscut angle significantly affect the superconducting critical current of YBCO films deposited on such substrates, leading to anisotropy of $J_c$ in the basal plane. For large miscut angles (4° and 8°) the anisotropy is mainly attributed to flux pinning effects generated by the presence of APB's. For lower miscut angles (0.4°), for which an isotropic $J_c$ would be expected, the observed anisotropy is related to YBCO lattice strain in the basal plane. In both cases (pinning or strain effects) the anisotropy is significant only at low temperatures and magnetic fields, where the effects of thermal activation are reduced. We find that APB's provide additional pinning (higher $J_c$) in the longitudinal direction, but these effects are washed out by applied magnetic fields in excess of 0.3 T. At the same time, the $J_c$ suppression along the transverse direction is significant only at zero field and for temperatures lower than 70 K. At higher temperature, and/or in fields above 0.1 T, the $J_c$ measured along the T direction is comparable to that of a regular, non-miscut YBCO film. We conclude that a random array of APB's eventually present in YBCO coated conductors is not likely to produce visible effects on the critical current, particularly considered that many wire applications require magnetic fields well in excess of 0.1 T and operational temperatures above 60 K.

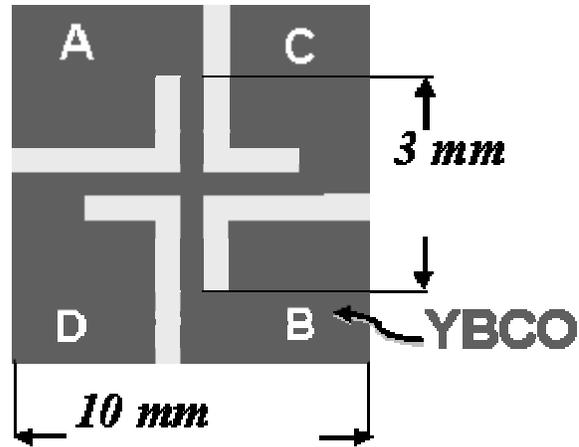

Fig. 1: Schematics of the cross conducting pattern used to measure electrical transport in L and T directions independently.

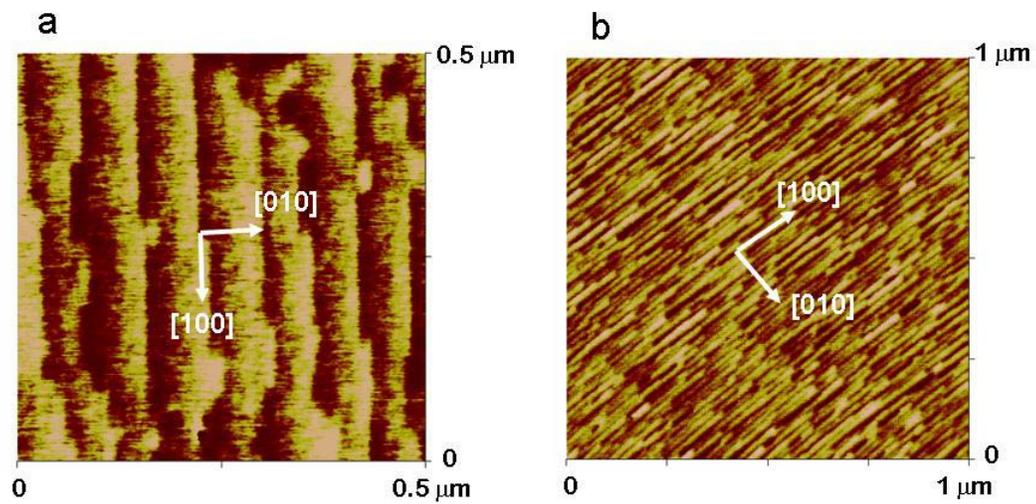

Fig. 2: (Color online) AFM micrographs of vicinal STO substrates annealed in pure $O_2$ at 1000 °C. The vicinal angle is 0.4° in a and 4° in b.

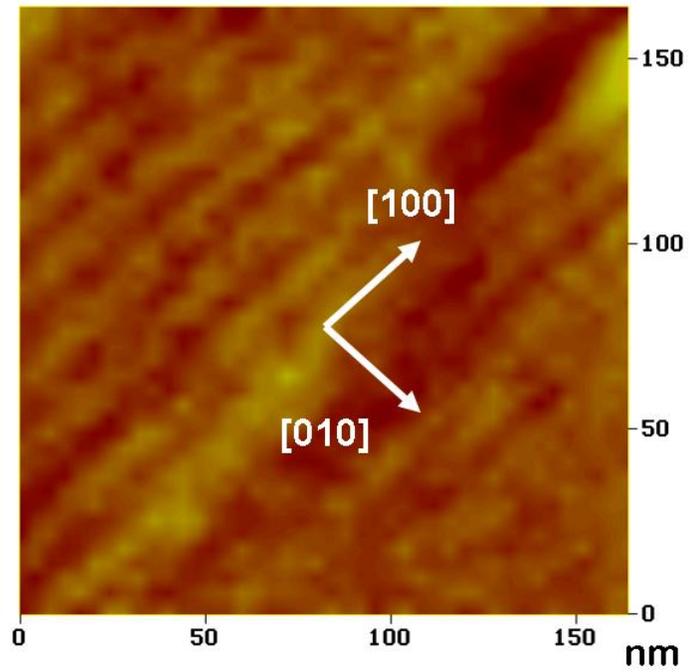

Fig. 3: (Color online) AFM micrograph of a YBCO film grown on a 8° vicinal STO substrate.

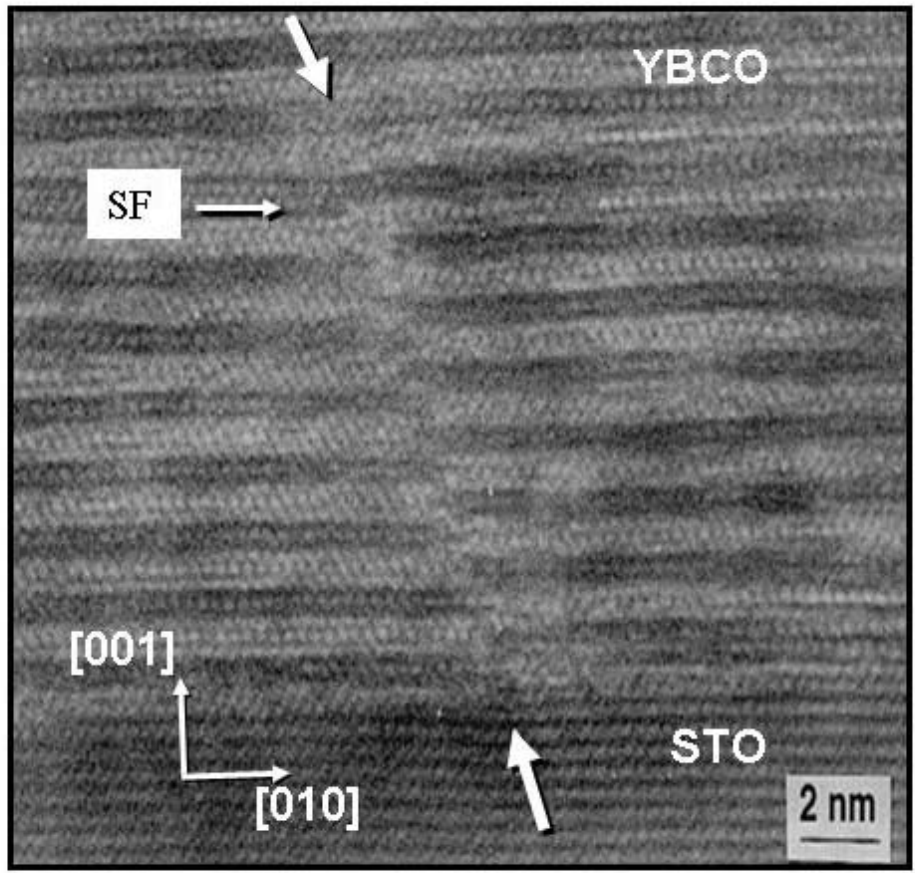

Fig. 4: High resolution TEM micrograph showing a cross section of YBCO on STO in proximity of the interface and along the transverse direction. A step in the STO substrate is seen as indicated by the bottom arrow. An antiphase boundary has formed in the YBCO film in correspondence with the substrate step and it extends until the proper stacking of YBCO planes is interrupted (as indicated by SF, or stacking fault) so that the adjacent YBCO planes now match. Originally appeared in Physica C 252 (1995) 125.

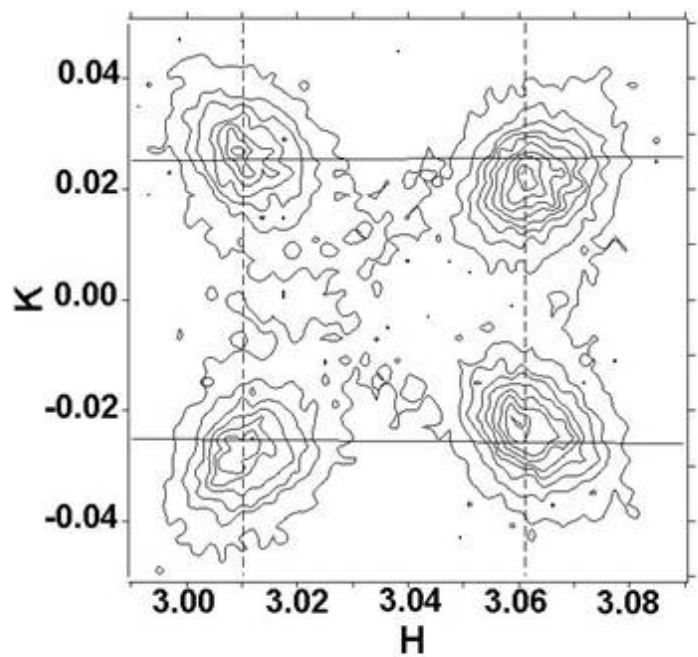

Fig. 5: reciprocal space x-ray map for the YBCO film grown on a 8° vicinal STO

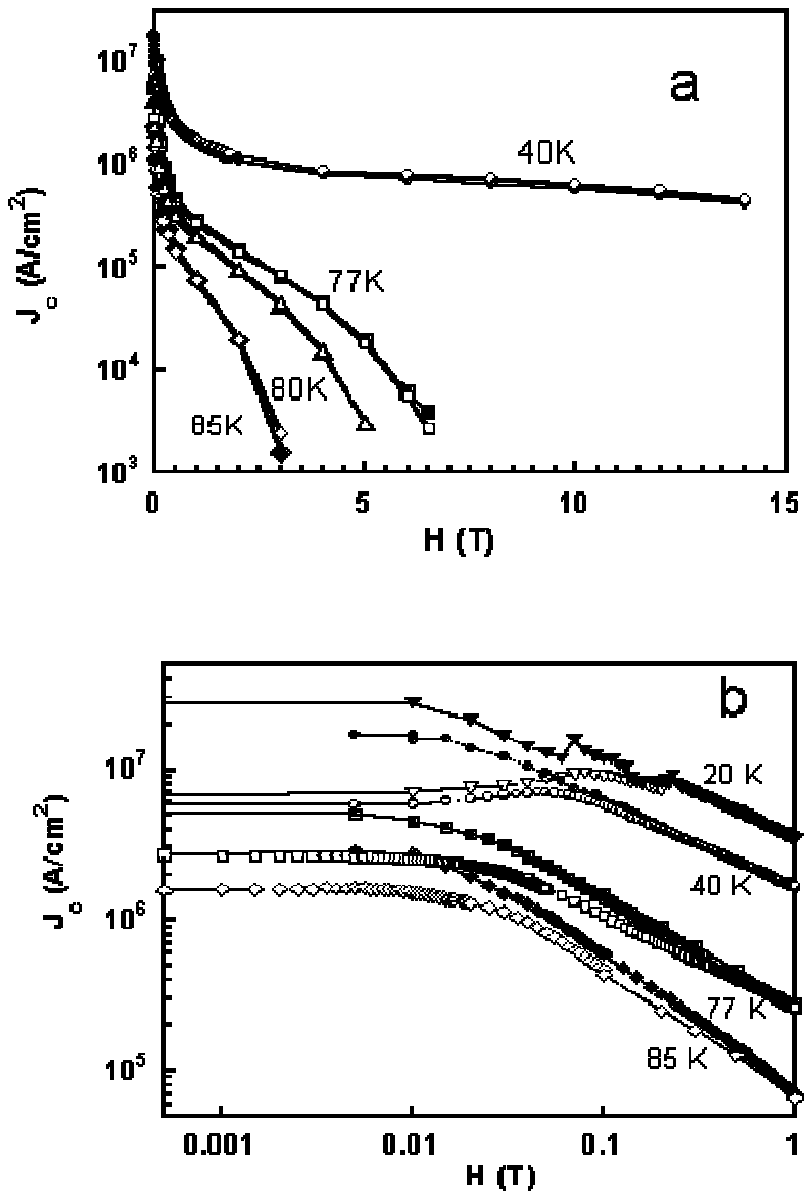

Fig. 6: $J_c$ dependence on applied magnetic field for the YBCO film on the 8° vicinal substrate (a) measured along L (filled symbols), and along T (empty symbols). Expanded view of the 40 K, 77 K, and 85 K data in the low field region with the addition of a data set at 20 K (b).

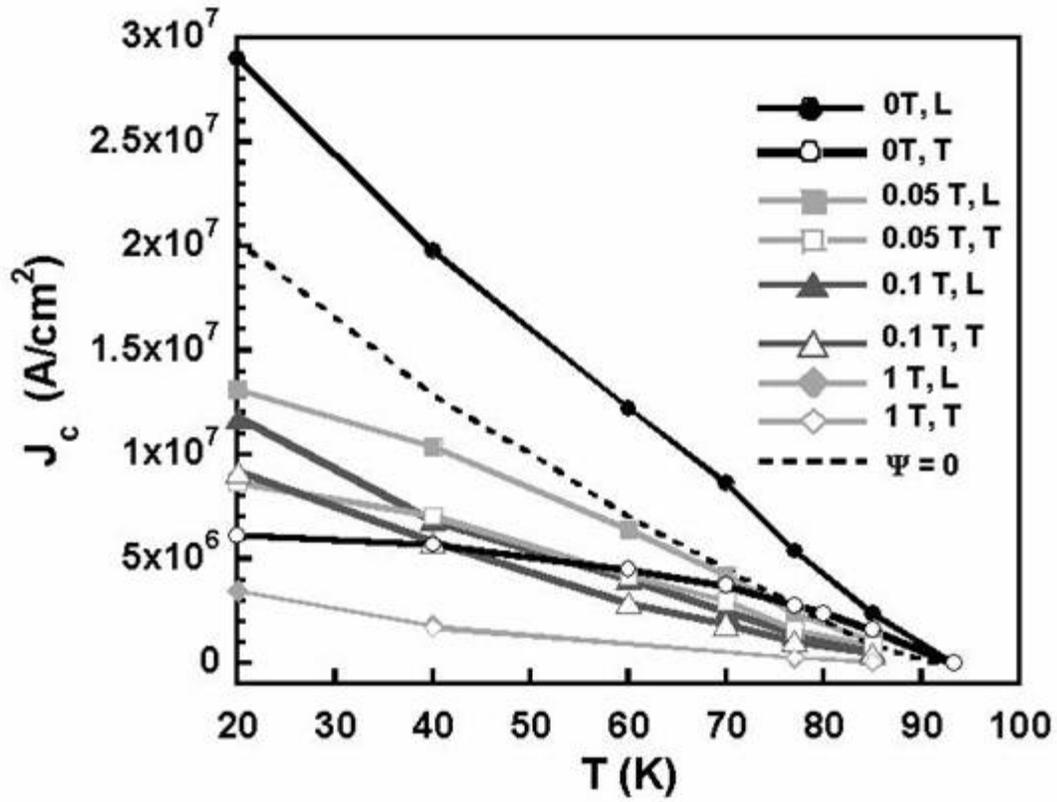

Fig. 7: Temperature dependence of $J_c$ for the L (filled symbols) and T (empty symbols) directions in the YBCO film grown on the 8° miscut substrate at different applied magnetic fields. The dashed line is the $J_c(T)$ curve measured in self field for a YBCO film grown on a flat surface.

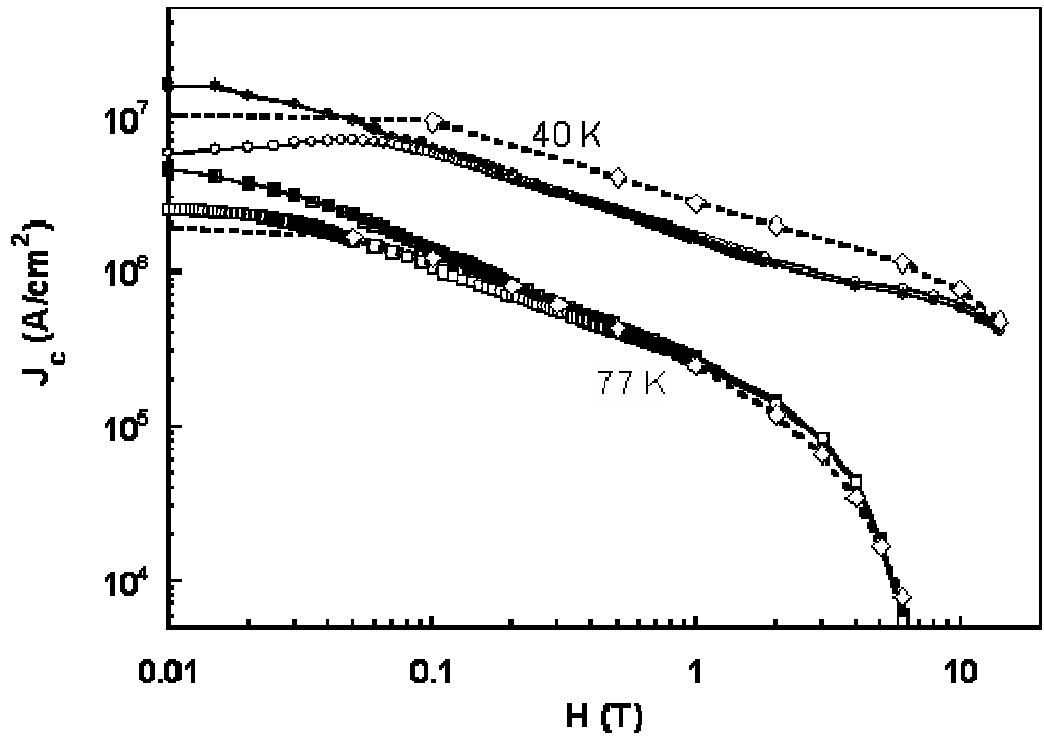

Fig. 8: Comparison between $J_c(H)$ curves for the YBCO film on flat STO (open diamonds with dashed line) and the YBCO film on 8° miscut STO at 77 K and 40 K (squares and circles). The empty symbols refer to the T direction, and the filled symbols to the L direction.

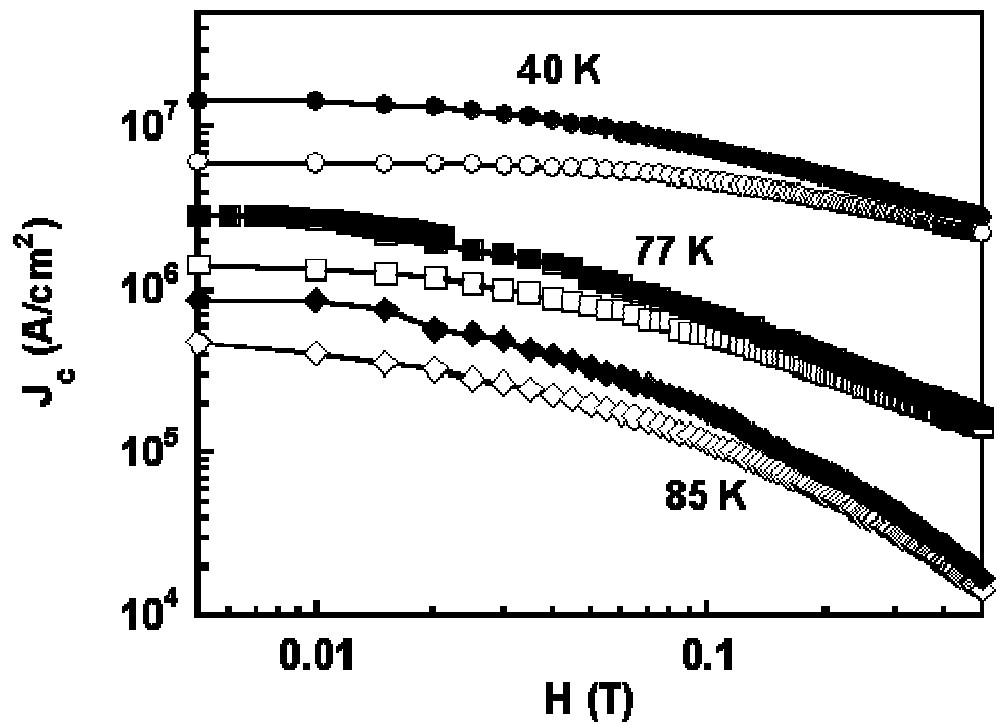

Fig. 9: $J_c$ dependence on applied magnetic field for an YBCO film grown on a 4° miscut STO. The empty symbols refer to the T direction and the filled symbols refer to the L direction.

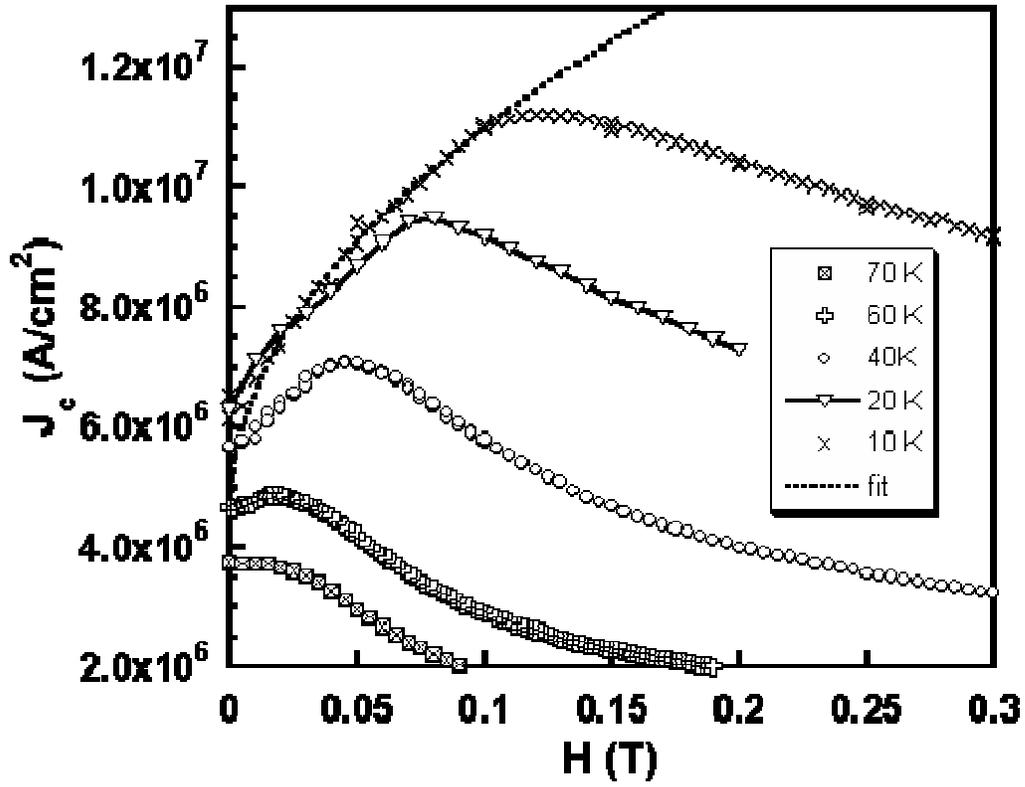

Fig. 10: $J_c$(H) dependence for small applied magnetic fields for the YBCO on 8° miscut STO along the T direction

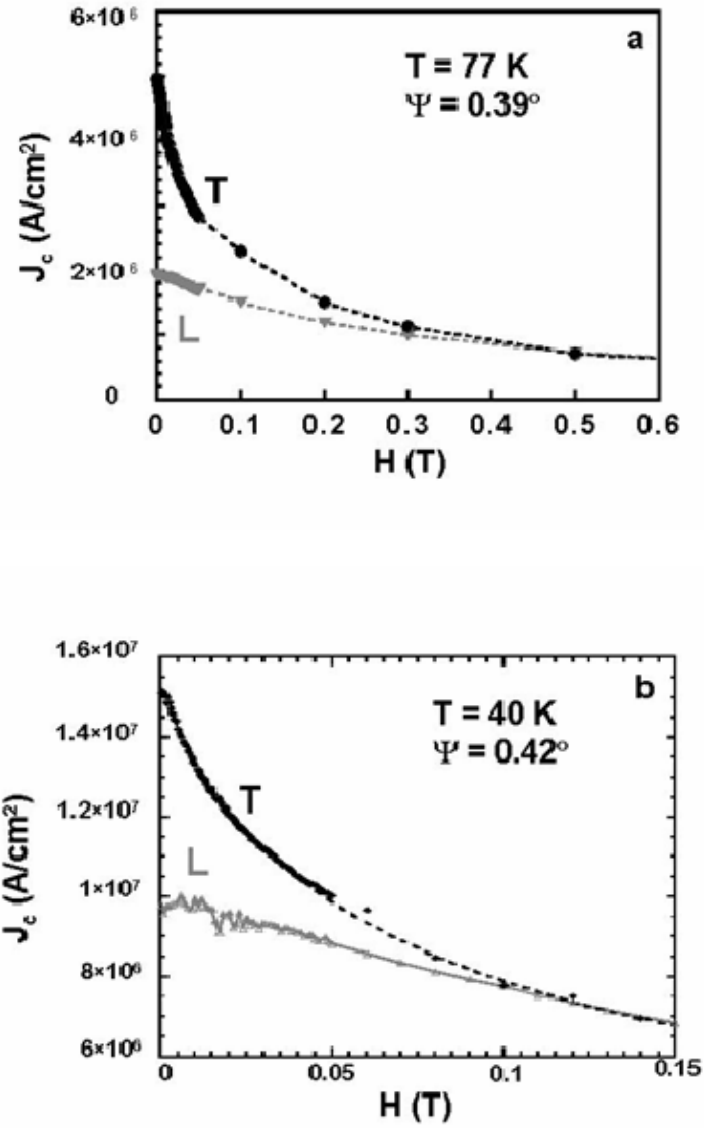

Fig. 11: $J_c$ vs. magnetic field measured in L and T directions for the YBCO film grown on the 0.39° miscut substrate in tables 1 and 2 (a), and for the YBCO film grown on the 0.42° miscut substrate in tables 1 and 2 (b).